\documentclass{article}
\usepackage{sao2,psfig}
\issue{2000, 49, 53-100}
\begin{document}
\markboth{Verkhodanov, Andernach \& Verkhodanova}
	 {Radio identification of decametric sources. I. Catalogue}
\title{Radio identification of decametric sources.\\ I. Catalogue}
\author{O.V. Verkhodanov\inst{a}
    \and H. Andernach\inst{b}
    \and N.V. Verkhodanova\inst{a}
}
\institute{\saoname
\and Depto.\,de Astronom\'\i a, Apdo.\ Postal 144, Univ.\ Guanajuato,
	     Guanajuato, Mexico}
\date{February 18, 2000}{May 3, 2000}
\maketitle

\begin{abstract}
We describe a method of plotting continuous spectra for radio sources
on the basis of entries of various source catalogues in comparatively
large error boxes around a given sky position.  Sources from the
UTR--2
catalogue (Braude et al., 1978, 1979, 1981, 1985, 1994), observed at decametric wavelengths
(10--25\,MHz) with an antenna beam of about 40$'$, were
cross-identified with entries from other radio catalogues at higher
frequencies.
Using the CATS database
we extracted all sources within 40$'$ around
UTR positions to find candidate identifications.  The spectrum for each source
was fitted with a set of curves using the least squares method to find
the best fit. We preferentially selected radio counterparts whose
radio spectrum extrapolated to low frequencies matched the UTR decametric 
flux densities, and whose coordinates
were close to the gravity centre of UTR positions. Among all 1822
sources in the UTR catalogue we found about 350 sources to be blends of 
two or more sources. As the most probable true coordinates of the radio
counterparts we used positions from the FIRST,
NVSS, TXS,
GB6, or
PMN catalogues.
Using low-frequency sources (26, 38, 85\,MHz) from CATS we checked the 
reliability of some of our identifications. We show examples of the above methods,
including raw and ``cleaned'' spectra.
Identifications with objects from optical, X-ray and infrared 
catalogues, as well as a subsample of UTR sources with ultra-steep 
spectrum are presented.
\end{abstract}

\section{Introduction}
The catalogue of 1822 radio sources obtained with the UTR telescope
near Kharkov (Braude et al., 1978, 1979, 1981, 1985, 1994,
{\tt http://www.ira.kharkov.ua/UTR2/})
covers about 30\% of the sky at six frequencies from 10 to 25~MHz,
and is currently the lowest-frequency catalogue of its size.
It provides an ideal basis to study the little known
optical identification content of sources selected at decametric
frequencies. The optical identification rate in the original 
version of the UTR-2 catalogue (UTR in what follows) is only 19\%.
Our goal is to identify all UTR sources with known radio sources.
This cross-identification yields both the radio continuum spectrum
and accurate coordinates for the radio counterparts, thus allowing
one to search for optical counterparts on the Digitized Sky Surveys.
The radio spectral characteristics may also be used to select 
certain samples, e.g. steep radio spectra and the lack of an 
optical identification tend to point to high-redshift radio galaxies.
 
The very large uncertainties of the UTR source positions 
($\sim$0.7$^{\circ}$) forced us to use an interactive process 
to derive radio source spectra.  In this process we use the radio sources
known from low- and intermediate-frequency catalogues as the most likely
candidates, which help us to discard the multitude of weaker sources in
the UTR error box provided e.g. by the recent and sensitive
NVSS (Condon et al., 1998) and FIRST (White et al., 1997) source catalogues.
All catalogue entries are extracted from the
catalogue collection combined in the CATS database (Verkhodanov et al.,
1997a).

    \begin{figure*}[!t]
    \centerline{
    \hbox{
    \psfig{figure=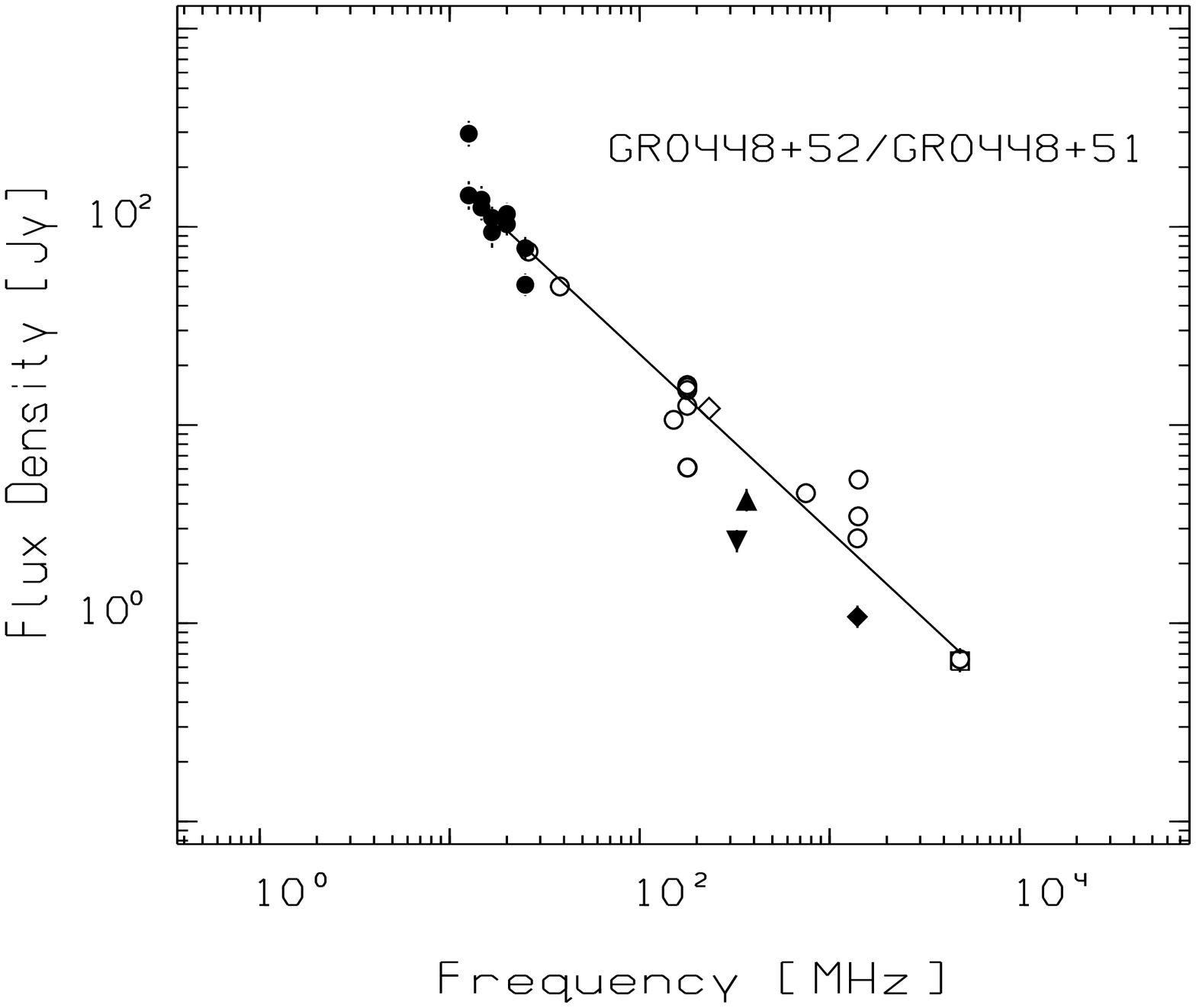,width=8cm,height=6cm,angle=-90,bbllx=10pt,bblly=100pt,bburx=530pt,bbury=740pt,clip=}
    \hspace*{0.7cm}
    \psfig{figure=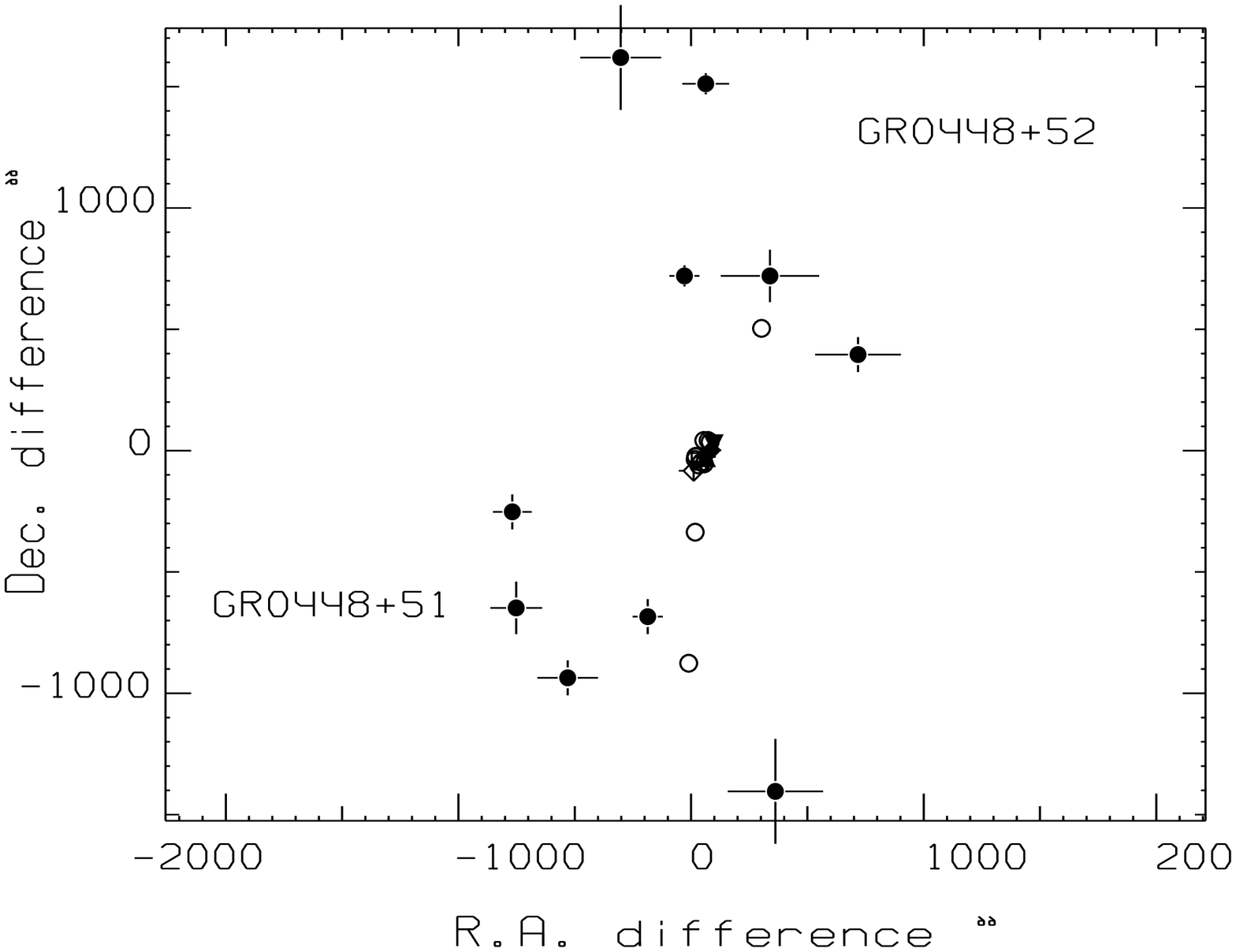,width=7cm,height=6cm,angle=-90,bbllx=10pt,bblly=100pt,bburx=520pt,bbury=730pt,clip=}
    }}
    \vspace*{-3ex}
    \caption{
 Example of a real source (4C+52.10) which has been observed as two
   independent UTR sources in adjacent observational strips.
 The left panel shows a spectrum, and the right panel shows the
   position of a confused source.
  Horizontal and vertical bars indicate the positional errors in
R.A. and DEC, respectively.
 }
    \end{figure*}

\begin{figure*}[!t]
\centerline{
\hbox{
\psfig{figure=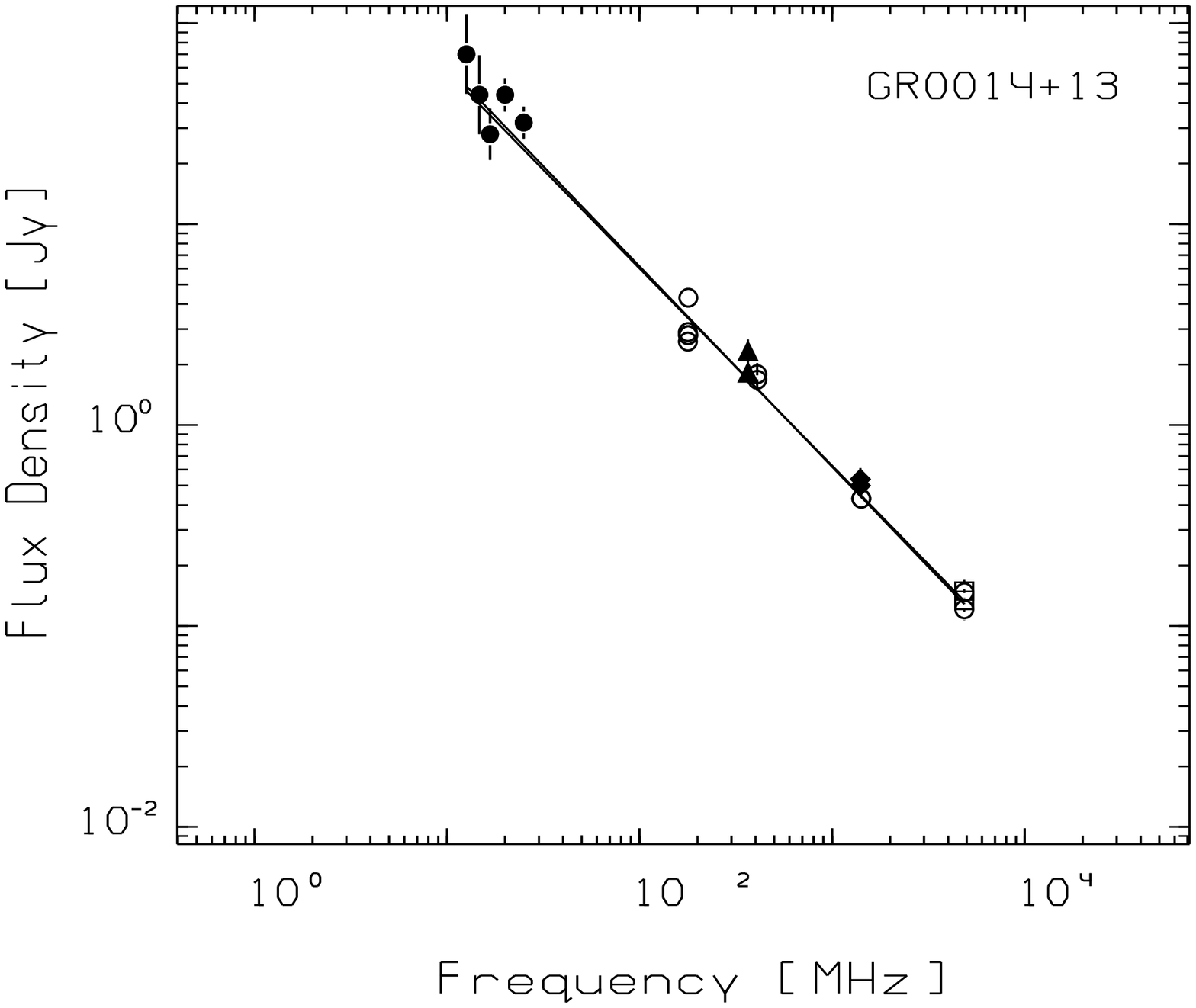,width=8cm,height=6cm,angle=-90,bbllx=10pt,bblly=100pt,bburx=530pt,bbury=740pt,clip=}
\hspace*{0.7cm}
\psfig{figure=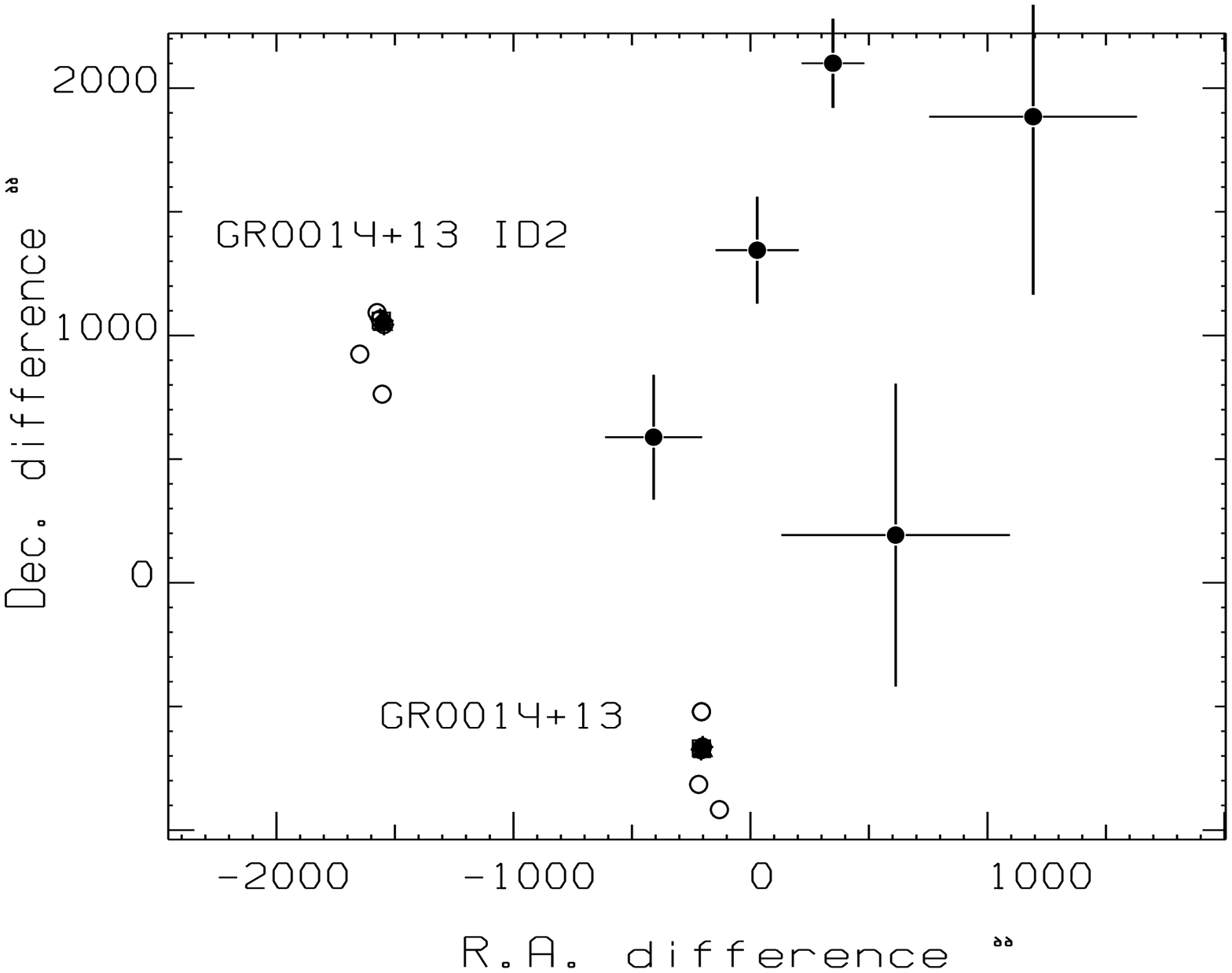,width=7cm,height=6cm,angle=-90,bbllx=10pt,bblly=100pt,bburx=520pt,bbury=730pt,clip=}
}}
\vspace*{-3ex}
\caption{
Left panel: spectra of two sources contributing to one UTR object.
Their contribution in the radio spectrum is practically
indistinguishable.  The right panel shows the relative positions of all
radio catalogue entries contributing to the spectrum.  The data have
already been cleaned from other irrelevant sources in the area.  Filled
circles correspond to the UTR catalogue entries, with horizontal and
vertical bars indicating the positional errors in R.A. and DEC,
respectively.  Two groups of non-UTR points (corresponding to 2
blending sources within the UTR beam) are visible at the lower and left
edge of the chart.
}
\end{figure*}
  The work on the revision of coordinates and spectra of the decametre
sources of the UTR catalogue was
initiated by the authors in 1996 and has been conducted until the present
time (Andernach et al., 1997; Verkhodanov et al., 1997b,c, 1998a,b;
Verkhodanova et al., 1998).
The publication of new radio source measurements implies a permanent
refinement of continuum radio spectra
bringing about minor changes in their parameters.
More recently we started to work on the identification of samples of the
UTR catalogue (Andernach et al., 2000; Verkhodanov et al., 2000), and the
requests of other interested radio astronomers, as well as
the authors of the UTR catalogue
motivated us to publish the procedures of identification and the resulting
list.


The main goal of the present paper is to derive precise
coordinates of the decametric sources of the UTR survey, which will further
be used for cross-identifications. One of the tasks was establishment
of cross-identification lists in the
form of files for each object and their inclusion in the CATS system.
For this
reason, we do not give here flux densities at different wavelengths,
except for
18 objects of the RC catalogue (Parijskij et al., 1991, 1992) expected
to be useful in further studies by the RATAN--600 group.
  All the information about the fluxes and references to
their original papers is
available from the CATS database (http://cats.sao.ru).

\section{Plotting of radio continuum spectra}
To prepare radio continuum spectra for decametric sources of the 
UTR catalogue (Braude et al., 1978, 1979, 1981, 1985, 1994), detected at 10, 12.6,
14.7, 16.7, 20, and 25~MHz, we first need to identify the sources 
with other known radio sources within their large error boxes 
(we used a box of 40$'\times\,$40$'$) drawn from the CATS database.
CATS provides a graphical interface which displays a ``radio spectrum''
for all sources found in the error box at various frequencies.
By human interaction the most deviant flux measurements in the spectrum
can be recognized as inappropriate counterparts, and are discarded
from the spectrum. This ``cleaning'' is achieved
with the programme {\it spg} (Verkhodanov, 1997).

The steps used in the identification of UTR sources 
are the following\,:
\begin{enumerate}
\item
    From the main radio catalogues of the CATS database
    (Verkhodanov et al., 1997a) we extract all entries in the search
    box of 40$'\times\,$40$'$, except for the recent and very sensitive
    NVSS and FIRST catalogues.
\item
    All objects with flux measurements at several frequencies are
    separated in the search box of 40$'\times\,$40$'$.
\item
    The spectrum of each object, excluding the UTR data points, is
    fitted with the best of several curves (see sect. 3) and extrapolated
    to the UTR frequencies.
\item
    Inside the search box we select counterparts by the
    following rules: \\[-4.5ex]
    \begin{enumerate}
    \item [(a)]
       the decametric flux densities, as extrapolated from the fitted spectra,
       should be close to the observed UTR fluxes;
    \item [(b)]
       positions of the radio counterparts should be close to the mean
       position as listed in the UTR catalogue.
    \end{enumerate}
    The resulting number of candidate identifications per UTR source
    ranges from 1 to 4 (see also Fig.~1).
    In the case of more than one counterpart,
    we consider that all counterparts satisfying the described criteria
    contribute to the UTR source flux, i.e. the UTR detection is a result 
    of blending of one or more independent sources. \\[1ex]

\item
    For further (e.g.\ optical) identification we defined ``best radio
    coordinates'' from the following catalogues (in order of decreasing
    priority): TXS (365~MHz), GB6 (4850~MHz), 87GB (4850~MHz), PMN (4850~MHz).
    It turns out that all UTR sources (with three exceptions, see below)
    have a counterpart from at least one of these catalogues.
\item
    If the identification area is poor of objects (e.g. at low declination,
    covered by only few radio surveys), and there are no sources detected
    simultaneously at several frequencies
    (i.e.\ no spectral fit is possible),
    then all objects within the box are retained for further study.
\item
    The best radio coordinates were then used for identification with NVSS or
    FIRST sources. The use of flux densities  from NVSS or FIRST usually improved
    the smoothness of the radio spectra.
\item
    Only if an NVSS or FIRST identification was found,
    the ``best radio positions''
    of item 5 were overwritten with the NVSS/FIRST position.
\item
    The ``best radio positions'' are used for identification of UTR objects 
    with optical object catalogues (e.g.\ from the APM scans of POSS or
    ESO/SERC
    surveys) or catalogues in other wavelength ranges.
\end{enumerate}

One of the problems of identification inside a wide antenna beam is
source confusion, i.e.\ when more than one strong source is located
within the UTR telescope beam and contributes to the UTR flux. As
mentioned in item (4) above, we have considered such sources and mark 
them as 'ID\#' in Table~1.  An example of one such source is shown in Fig.~2.

\noindent
To check the reliability of the derived spectra we used the 
low frequency catalogues
6C, 7C at 151~MHz
(Baldwin et al., 1985; Hales et al., 1988, 1990, 1991, 1993a, 1993b;
McGilchrist et al., 1990),
3C, 4C at 178~MHz
(Edge et al., 1959; Bennet, 1961;  Pilkington and Scott, 1965),
and other catalogues included in Dixon's Master List, like
CL (Viner \& Erickson, 1975, 26~MHz), WKB (Williams et al.,
1966, 38~MHz), MSH (Mills et al., 1958, 1960, 1961, 85~MHz).
  For the region of the sky where these surveys overlap with the area covered
by the UTR catalogue, we find that the low frequency fluxes from the former
surveys match our spectral fits for UTR sources, and that source coordinates 
from these surveys are close to those derived by our method.
This confirms the reliability of our methods of source separation.

\section{The catalogue of counterparts}
The resulting catalogue of 2316 radio counterparts, including all
multiple identifications,
is given in Table~1.
In the columns of this table we give the original (B1950-based) UTR 
source name, the right ascension and declination
(B1950 and J2000.0) of the best radio position,
Galactic latitude, spectral indices of the best-fitting spectrum
 at 365, 1400 and 5000 MHz,
best-fitting radio spectral parameters, 
presence of known optical, infrared or X-ray emission (marked as ``em.''
in Column 11),
and other names.
Column 1 lists a sequence number for each radio identification.
If there is more than one radio identification we assign
sequence numbers ID2, ID3, etc. in column 2 of Table~1.
However, we omit the ``ID1'' in all cases.
Column 3 lists a sequence number of another radio identification
if the latter significantly contributes to the UTR flux, or marks
confused sources with a symbol 'c'.
The spectrum is parameterized by the formula $ lg\,S(\nu) = A + Bx + C f(x)$,
where $S$ is the flux density
in Jy, $x$ is the logarithm of the frequency $\nu$ in MHz, and
$f(x)$ is one of
the following functions: $exp(-x)$, $exp(x)$, or $x^2$.
Identifications with known X-ray sources are indicated with an ``X''
in column (11) of Table~1, while infrared (``I'') and optical (``O'')
identifications are further detailed in separate Tables 4 and 5.
Column 12 lists the names of sources from the known radio catalogues:
3C (Edge et al., 1959; Bennett, 1961),
4C (Pilkington and Scott, 1965) and
PKS (Otrupcek and  Wright, 1990),
with which the UTR sources are identified.

Only for three sources we were unable to find identifications 
among the catalogues described above:
GR\,0801$-$11, GR\,0930$-$00, and GR\,1040$-$02.

There are not enough data to be sure of the identifications of the
sources
GR\,0520$-$08,
GR\,0537$-$00, GR\,0629+02, and GR\,2345+03.
For these sources we quote a less accurate position and mark the position
with a question mark.

We worked with both existing versions of the UTR catalogue:
the printed version  (Braude et al., 1978, 1979, 1981, 1985, 1994)
and the more recent electronic one ({\tt http://www.ira.kharkov.ua/UTR2/}).
We included UTR sources of all reliability levels (A,B,C) as given in
the UTR catalogue.
There are 64 sources in the electronic
version which are not in the printed version, and 4 sources vice versa.
Some of the sources are listed with different names in the printed and
electronic versions.

In Table~1 we kept the names from the printed catalogue version for the
following sources
(new names from the electronic version are given in brackets):
GR\,0224+03 (GR\,0227+03),
GR\,0307+17 (GR\,0307+16),
GR\,0411+14 (GR\,0411+13),
GR\,0919+55 (GR\,0918+55),
GR\,0929+07 (GR\,0930+07),
GR\,1039+03 (GR\,1039+02),
GR\,1142+00 (GR\,1142$-$00),
GR\,1538+01 (GR\,1539+01),
GR\,1547+03 (GR\,1548+03).

\medskip
Table~1 also includes the following sources,
which are present in the printed version,
but are absent in electronic one\,:
  GR\,1915+56,
  GR\,2355+19,
  GR\,2355$-$02,
  GR\,2358+08.

\setcounter{table}{0}
\begin{onecolumn}
\tiny
\mbox{\hspace*{15cm}}\\
\topcaption{
Radio Identifications for all UTR catalogue sources
}
\tablefirsthead{
  \hline \multicolumn{1}{|c}{N}        
    & \multicolumn{1}{|c}{name\hspace{5mm}conf.}
    & \multicolumn{1}{|c}{RA+Dec(B1950)}   
    & \multicolumn{1}{|c}{RA+Dec(J2000)}   
    & \multicolumn{1}{|c}{$b$}             
    & \multicolumn{1}{|c}{$\alpha_{365}$ $\alpha_{1400}$ $\alpha_{5000}$}  
    & \multicolumn{1}{|c}{spectrum}        
    & \multicolumn{1}{|c}{em.}             
    & \multicolumn{1}{|c|}{radio names}\\   
      \multicolumn{1}{|c}{ 1 }         
    & \multicolumn{1}{|c}{ 2\hspace*{12mm}3}
    & \multicolumn{1}{|c}{ 4 }         
    & \multicolumn{1}{|c}{ 5 }         
    & \multicolumn{1}{|c}{ 6 }         
    & \multicolumn{1}{|c}{ 7\hspace*{5mm}8\hspace*{5mm}9 }
    & \multicolumn{1}{|c}{ 10 }        
    & \multicolumn{1}{|c|}{ 11 }       
    & \multicolumn{1}{|c|}{ 12 } \\    
      \multicolumn{1}{|c}{\it N}                           
    & \multicolumn{1}{|c}{\hspace*{12mm}\it~N}              
    & \multicolumn{1}{|c}{$^h~^m~^s$+\degr~\arcmin~\arcsec}
    & \multicolumn{1}{|c}{$^h~^m~^s$+\degr~\arcmin~\arcsec}
    & \multicolumn{1}{|c}{ \degr }                         
    & \multicolumn{1}{|c}{ }                               
    & \multicolumn{1}{|c}{ }                               
    & \multicolumn{1}{|c|}{}                               
    & \multicolumn{1}{|c|}{} \\ \hline                     
	     }
\tablehead {\hline
      \multicolumn{1}{|c}{ 1 }
    & \multicolumn{1}{|c}{ 2\hspace*{12mm}3}
    & \multicolumn{1}{|c}{ 4 }
    & \multicolumn{1}{|c}{ 5 }
    & \multicolumn{1}{|c}{ 6 }
    & \multicolumn{1}{|c}{ 7\hspace*{5mm}8\hspace*{5mm}9 }
    & \multicolumn{1}{|c}{ 10}
    & \multicolumn{1}{|c|}{ 11 }
    & \multicolumn{1}{|c|}{ 12 }     \\
      \multicolumn{1}{|c}{ }                               
    & \multicolumn{1}{|c}{ \hspace*{12mm}\it~N}
    & \multicolumn{1}{|c}{$^h~^m~^s$+\degr~\arcmin~\arcsec}
    & \multicolumn{1}{|c}{$^h~^m~^s$+\degr~\arcmin~\arcsec}
    & \multicolumn{1}{|c}{ \degr }                         
    & \multicolumn{1}{|c}{$\alpha_{365}$ $\alpha_{1400}$ $\alpha_{5000}$}  
    & \multicolumn{1}{|c}{}                                
    & \multicolumn{1}{|c|}{}                               
    & \multicolumn{1}{|c|}{} \\ \hline                     
	 }
\tabletail{\hline}

\begin{center}

\end{center}
\end{onecolumn}

\section{Spectra}
Characteristics of the main catalogues used in the identification
process are given in Table 2.

\setcounter{table}{1}
\begin{table}[!h]
{\small
\caption{Characteristics of basic catalogues used in radio identification}
\begin{center}
\begin{tabular}{|lrcr|}
\hline
Name   & Freq. &  HPBW($'$)   & S$_{lim}$(mJy) \\
\hline
6C     &    151 &  4.2       & $\sim$200 \\
7C     &    151 &  1.2       &     80    \\
MY     &    232 &  3.8       &$\sim$100  \\
TXS    &    365 &$\sim$0.1   &$\sim$200  \\
B3     &    408 &3$\times$5  &  100      \\
WB92   &   1400 &10$\times$11&  150      \\
87GB   &   4850 &  3.7       &   25      \\
GB6    &   4850 &  3.7       &   15      \\
PMN    &   4850 &  4.2       &   30      \\
MSL    &   misc.& misc.      & misc.     \\
\hline
\end{tabular}
\end{center}
}
\end{table}

In Table 2 the following references are provided for acronyms:
\begin{tabbing}
6C  \hspace*{0.5cm}  \quad\= Hales et al., 1993a, b,  \\
7C    \>      McGilchrist et al., 1990,    \\
87GB  \>      Gregory et al., 1991,        \\
B3    \>      Ficarra et al., 1985,        \\
GB6   \>      Gregory et al., 1996,        \\
MSL   \>      Dixon, 1970,                 \\
MY    \>      Zhang et al., 1997,          \\
PMN   \>      Wright et al.,  1996,        \\
TXS   \>      Douglas et al., 1996,        \\
WB92  \>      White et al., 1992.
\end{tabbing}
We cleaned spectra with the programme {\it spg} (Verkhodanov, 1997)
using the methods also described
in the paper by Verkhodanov et al. (2000).
For the radio-identified UTR objects we derived spectral
indices at three frequencies from the spectra of best fitting.
Distributions of these spectral indices at 85, 365 and 1400 MHz are shown
in Fig.\,3.
The fraction of flat spectra clearly increases with frequency,
consistent with the presence of a large number of concave spectra.
The distribution of radio spectra among the different spectral types
is shown in Table\,3 for all 2306 radio counterparts to UTR sources.

\begin{table}[!h]
\caption{Distribution of spectral types}
\begin{center}
\begin{tabular}{|llrr|}
\hline
Spectral class & Fitting function & Number & \% \\
\hline
Straight spectra    & $+A+Bx                  $ &  898   &  39 \\
Convex   ($C^{+}$)  & $+A{\pm}~Bx-Cx^2         $ &  184   &   8 \\
Concave  ($C^{-}$)  & $+A-Bx+Cx^2             $ & 1147   &  50 \\
		    & ${\pm}A{\pm}~Bx+Ce^{-x}  $ &   77   &   3 \\
\hline
\end{tabular}
\end{center}
\end{table}

\begin{figure*}[!t]
\centerline{
\hbox{
\psfig{figure=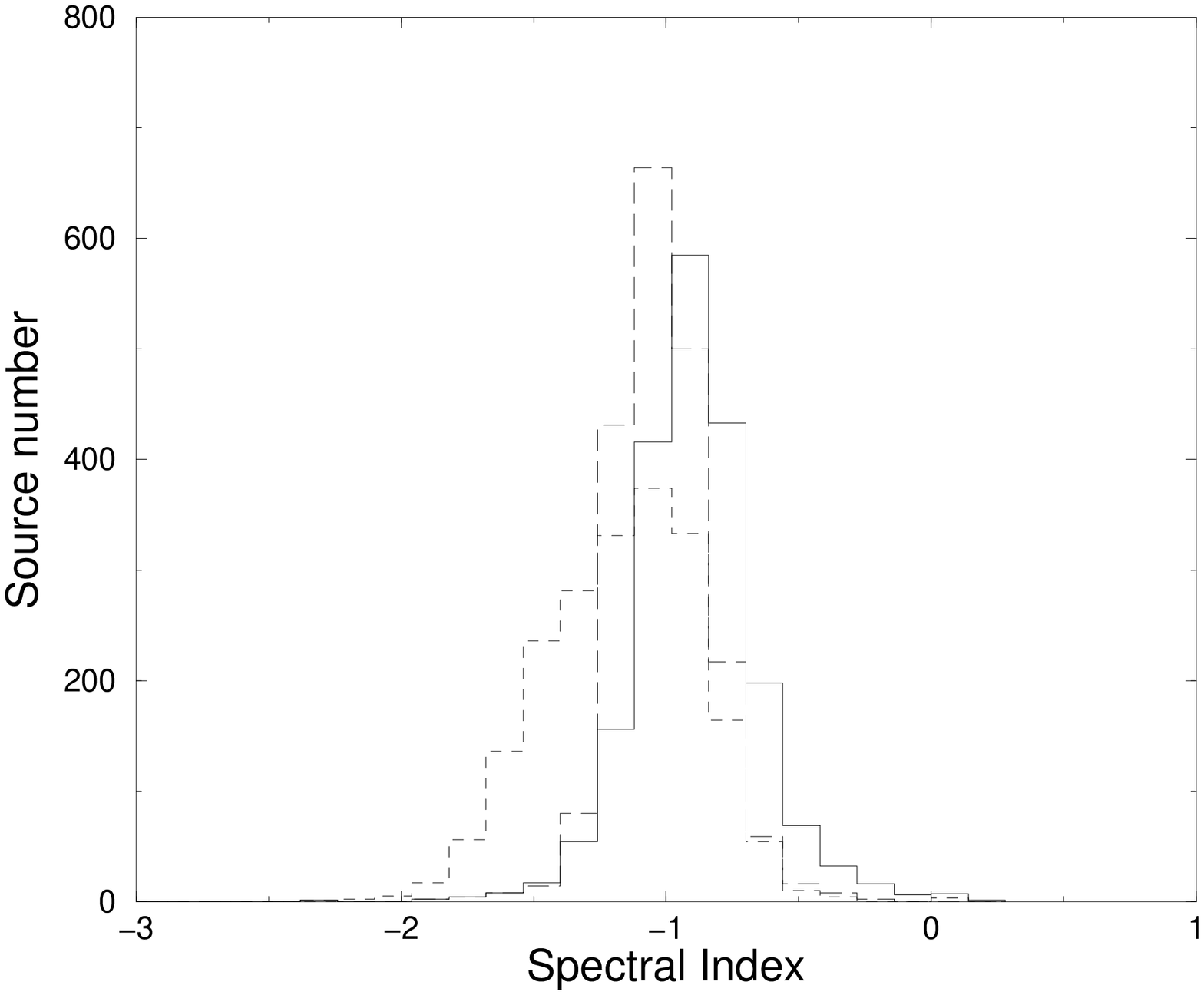,width=8cm,height=6cm,angle=-90}
\hspace*{0.5cm}
\psfig{figure=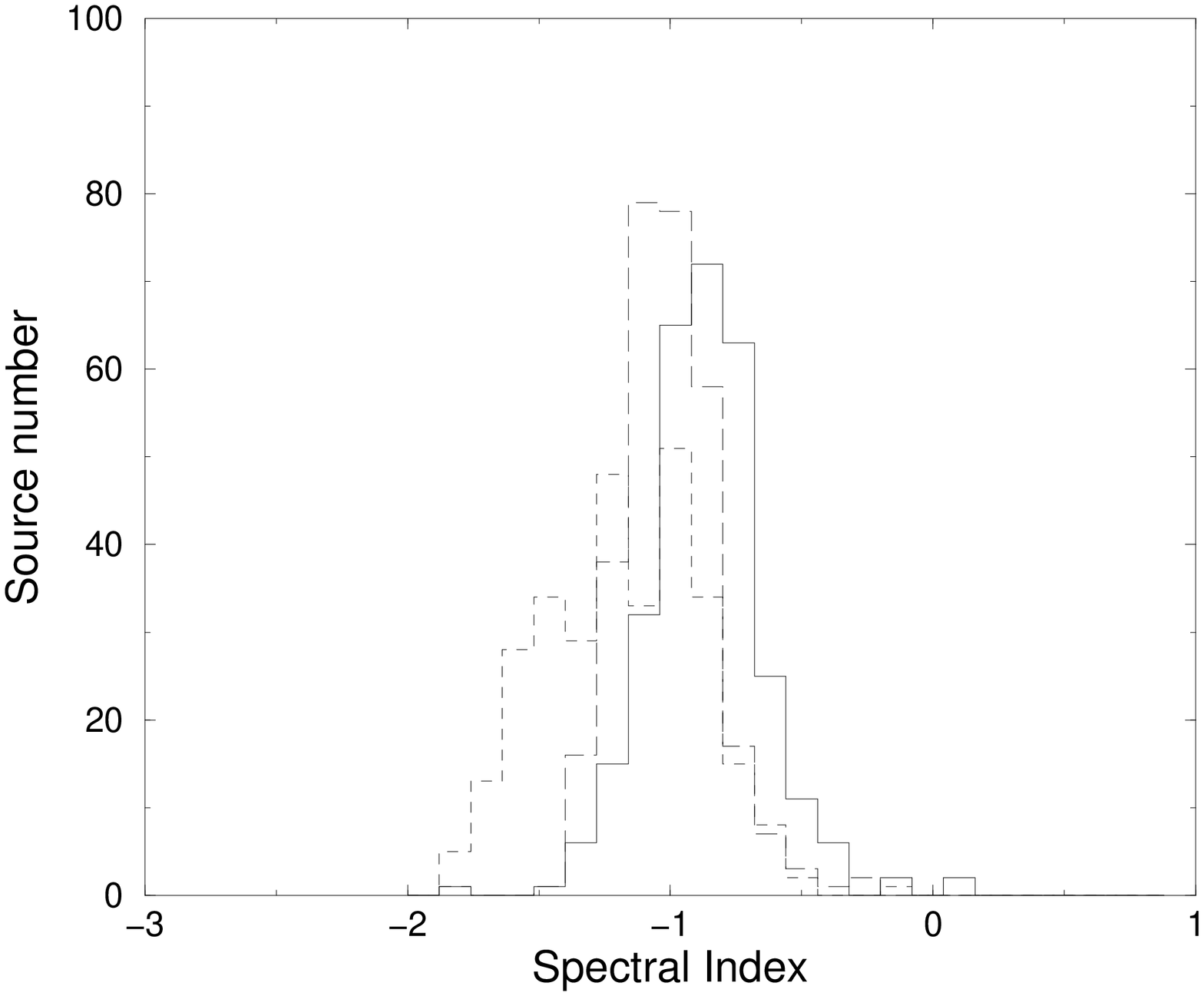,width=8.5cm,height=6cm,angle=-90}
}}
\caption{
Distributions of spectral indices of radio-identified sources
at frequencies 80\,MHz (short-dashed), 365\,MHz (long-dashed) and
1400\,MHz (solid lines) for  2006 sources with $|b|>10^\circ$ (left)
and for 301 sources near the Galactic plane ($|b|\le10^\circ$).}
\end{figure*}


From the best-fitting spectra of UTR sources we calculated spectral
indices at three frequencies (see Table~1 and Fig. 3).  The behaviour
of integrated spectra of radio sources in the decametre range has been
investigated, for instance, by Sokolov (1996), who also considered the
spectra of UTR sources, though without account for blending,
i.e.\ based on a single radio identification per UTR source.  The fact
that for the radio-identified UTR sources the spectral index
distribution shows steeper spectra at lower frequency is likely to be
due to two effects. One is the blending of various unrelated sources at
decametric frequencies, and the other is the fact that the UTR
instrument, working at low frequencies and low angular resolution, is
more sensitive to extended regions of radio galaxies (``lobes'') which
tend to have a steeper spectrum than the overall source (Miley et al.,
1980).

\section {Cross-identification with non-radio source catalogues}

We used the ``precise'' coordinates of Table~1 for cross--identifications
with optical and mixed catalogues like VV8AG, MCG, PGC, Dixon's Master
Source List (MSL) and other non-radio catalogues of the CATS database.
The result of cross-identification within a circle of $10\arcsec$ is given
in Table~4.
Note that our data tables are based only on catalogue information within
CATS, but do not include information contained in NED or SIMBAD.
We plan to improve our identifications in future by an additional
use of these databases.

{
\tiny
\topcaption {Result of search for optical counterparts
within a radius of $10\arcsec$ in the CATS optical catalogues}
\tablefirsthead{
  \hline \multicolumn{1}{|c}{Name}                  
    & \multicolumn{1}{|c|}{Optical counterparts} \\  
  \hline
}
\tablehead {
  \hline
      \multicolumn{1}{|c}{Name}                      
    & \multicolumn{1}{|c|}{Optical counterparts} \\  
  \hline
}
\tabletail{\hline}
\begin{center}
\begin{supertabular}{|ll|}
\hline
 GR0002$+$12       & VV83 0002$+$12                          \\
 GR0003$+$15       & MSLO BC 4C15.01 (16.4),  QSO2 (15),     \\
		   &      VV8Q PKS 0003$+$15                 \\
 GR0003$-$00       & MSLO BC 3C 2   (19.3)                   \\
 GR0004$+$05       & VV83  (19.5)                            \\
 GR0004$-$05       & VV83  NRAO 7 (19)                       \\
 GR0010$+$00       & VV83  PKS 0010$+$00                     \\
 GR0010$+$05       & VV83  OTL 0011$+$054  (16.5)            \\
\hline
\end{supertabular}
\end{center}
}

Table 4 lists information on one or more optical counterparts 
for 575 different UTR sources.
This makes 32\% of the entire UTR catalogue, as compared
to 19\% both in the printed and electronic version of the UTR catalogue.
Table~4 lists 635 different radio counterparts from Table~1, or 
27\% of all 2316 radio counterparts. Note that for a small percentage 
of sources, especially very extended sources near the Galactic plane,
like SNRs and H\,II regions, discrete source catalogues tend to be
affected by dynamic range limitation, and the positions listed in Table~1
for these sources may be misleading. One such example is GR\,1901$+$05
which corresponds to the very extended SNR 3C\,396. For such sources
one needs to extend the optical search radius, study the large-scale
radio emission of the region (typically from single-dish maps of the
Galactic Plane) in order to pick up the correct optical identification.
We hope to do this in future  for more detailed analyses.

\begin{table*}[!t]
{\small
\caption{UTR sources with an infrared counterpart within a radius of
$60\arcsec$}
\vspace{0.1cm}
\begin{center}
\begin{tabular}{|ll|}
\hline
GR0238$+$58      & ISSC 023715$+$592255, ISSC 023715$+$592255, HII\_H 02381$+$5923 \\
GR0240$-$00      & IPSC 02401$-$0013, IFSC F02401$-$0013, ISSC 024018$-$001254     \\
GR0316$+$41      & IPSC 03164$+$4119, IFSC F03164$+$4119, ISSC 031551$+$412220     \\
GR0327$+$55      & IPSC 03261$+$5516                                               \\
GR0405$+$03      & IPSC 04047$+$0332                                               \\
GR0411$+$11      & ISSC 041037$+$114453                                            \\
GR0430$+$05      & IPSC 04305$+$0514, IFSC F04305$+$0514                           \\
GR0438$+$50      & IPSC 04368$+$5021, ISSC 043550$+$501908                         \\
GR0545$+$46      & IPSC 05457$+$4635                                               \\
GR0629$+$02      & IPSC 06306$+$0248                                               \\
GR0640$+$52      & IPSC 06401$+$5320, IFSC F06401$+$5320                           \\
GR0653$+$58      & IFSC F06540$+$5847                                              \\
GR0704$-$12      & IPSC 07029$-$1215, ISSC 070359$-$121320, HII\_I 07029$-$1215    \\
GR0707$+$47      & IPSC 07075$+$4715, IFSC F07075$+$4715                           \\
GR0723$-$09      & IPSC 07225$-$0933                                               \\
GR0807$-$10      & IPSC 08065$-$1018, IFSC F08064$-$1018                           \\
GR1000$+$55      & IPSC 09585$+$5555, IFSC F09585$+$5555, ISSC 095816$+$560631,    \\
		 & IRSSS X0958$+$559,                                              \\
GR1126$+$58      & IPSC 11257$+$5850, IFSC F11257$+$5850, ISSC 112544$+$584327     \\
GR1219$+$50(ID2) & IFSC F12163$+$5042                                              \\
GR1226$+$02      & IPSC 12265$+$0219, IFSC  F12265$+$0219, ISSC 122623$+$022029    \\
GR1228$+$12      & IPSC 12282$+$1240, IFSC F12282$+$1240                           \\
GR1254$-$05      & IFSC F12535$-$0530                                              \\
GR1319$+$42      & IFSC F13190$+$4251                                              \\
GR1320$+$43(ID2) & IFSC F13190$+$4251                                              \\
GR1328$+$47      & IPSC 13277$+$4727, IFSC F13277$+$4727, ISSC 132742$+$473356,    \\
		 & IRSSS X1327$+$474                                                 \\
GR1429$+$57(ID2) & IPSC 14306$+$5808, IFSC F14306$+$5808, IRSSS X1430$+$581        \\
GR1507$+$53      & IPSC 15080$+$5329                                               \\
GR1510$+$49      & IPSC 15099$+$5005                                               \\
GR1600$+$02      & IPSC 15599$+$0206, IRSSS B1600$+$0205                           \\
GR1650$+$02(ID2) & IFSC F16504$+$0228, IPSC  16504$+$0228                          \\
GR1812$+$42      & IPSC 18146$+$4238, IFSC F18146$+$4238                           \\
GR1901$+$05      & HII\_I 19008$+$0530, IPSC 19008$+$0530, HII\_H 19008$+$0530     \\
GR1921$+$16(ID2) & IPSC 19229$+$1614                                               \\
GR2025$+$15      & IPSC 20245$+$1527, IFSC  F20245$+$1527                          \\
GR2221$-$02      & IFSC F22212$-$0221, ISSC  222140$-$022325                       \\
GR2224$-$05      & IFSC F22231$-$0512, IPSC  22231$-$0512, ISSC  222300$-$051409   \\
GR2251$+$15      & ISSC 225156$+$155037                                            \\
GR2314$+$03      & IFSC F23140$+$0348, IPSC  23140$+$0348                          \\
GR2349$-$01      & IFSC F23493$-$0126                                              \\
\hline
\end{tabular}
\end{center}
}
\end{table*}

\begin{figure*}[!t]
\centerline{\psfig{figure=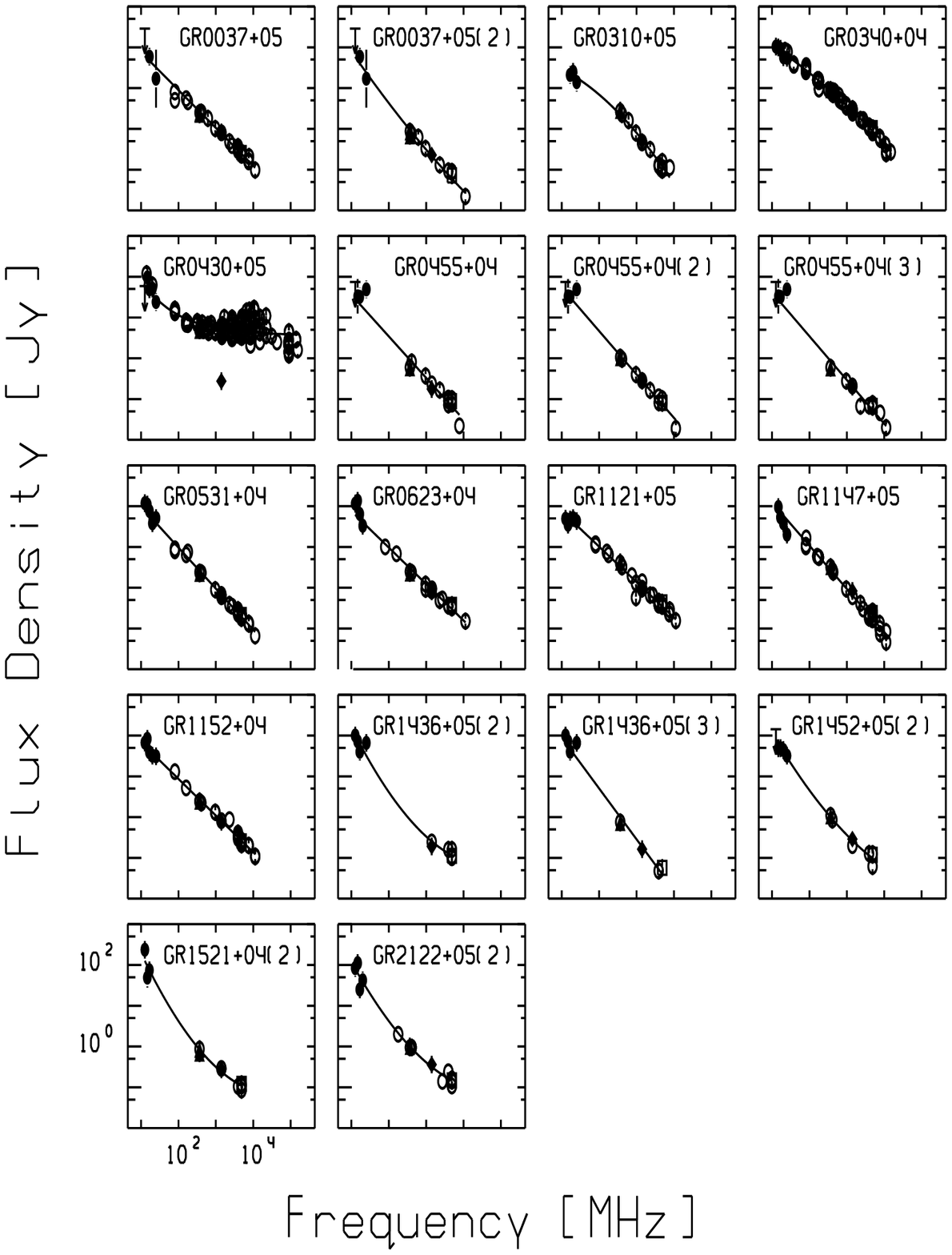,width=15cm}}
\vspace*{-6ex}
\caption{Radio spectra of RC objects with decametric data.
UTR points are shown with
filled ellipses, Texas data are shown with filled triangles and
NVSS data are shown with diamonds.}
\end{figure*}

In the first column of the table are given the names of the UTR objects
with their appended ID number from Table~1,
in the second column are the names of the catalogues and the objects
for which information is available in these catalogues accessable in CATS.
If stellar magnitudes are indicated
in the catalogues, they are given
in brackets. Unfortunately, the publications presented are compilations, and
data on the filters used to obtain stellar magnitudes are lacking. However,
they can be found in the original references quoted in the
individual compilation. This paper does
not seek to give complete information on the presence of optical
identifications,
we have just made cross-identification with the catalogues available in CATS.
Three points should be borne in mind:
\begin{enumerate}
\item
      The presence of a name in column~2 of Table~4 does not imply the
      existence
      of a candidate optical identification. In some cases our entry refers
      to an ``empty field'' (``EF''), or to a lower limit of the stellar
      magnitude.
\item Not all the lists of optical objects, including the ones identified with
      radio sources, are available in CATS.
\item 138 sources of the 3C catalogue are identified with the UTR
      objects, but for all 3C sources optical identifications presently exist.
\end{enumerate}
In column 2 of Table 4 the following references are tabulated:
\begin{tabbing}
3C \hspace*{2cm}  \quad\=  Spinrad et al., 1985,      \\
DENIS             \>       Vauglin et al., 1999,      \\
GAvZ1             \>       Weinberger et al., 1995,   \\
GCS               \>       da Costa et al., 1998,     \\
LBQS6             \>       Hewett et al., 1995,       \\
LCRS              \>       Shectman et al., 1996,     \\
MKN               \>       Markarian et al., 1989,    \\
MCG               \>       Kogashvili, 1982,          \\
MRC               \>       Large et al., 1991,        \\
MSLO              \>       Dixon and Sonneborn, 1980, \\
NBG               \>       Tully, 1988,               \\
PGC               \>       Paturel et al., 1989,      \\
QSO1              \>       Hewitt \& Burbidge, 1991,  \\
QSO2              \>       Hewitt \& Burbidge, 1993,  \\
UCGC              \>       Cotton et al., 1999,       \\
VV83              \>       V\a'eron-Cetty \& V\a'eron, 1983, \\
VV8AG \& VV8Q     \>       V\a'eron-Cetty \& V\a'eron, 1998.
\end{tabbing}

Given that positional errors for IR sources are significantly larger than
for optical sources, we have chosen a search radius of $60\arcsec$ for the
radio --- IR correlation. All sources within this distance from the positions
quoted in Table~1 are listed in Table~5.

In Table 5 the following references are provided:
\begin{tabbing}
HII\_H  \quad\=  Hughes \& MacLeod, 1989,                         \\
HII\_I  \>       Codella et al., 1994                             \\
IFSC    \>       IRAS Faint source catalogue by Moshir,            \\
	\>       1989                                              \\
IPSC    \>       IRAS Point source catalogue by IRAS group,        \\
	\>       1987                                              \\
IRSSS   \>       IRAS Small Scale Structure Catalog by             \\
	\>       Helou and Walker, 1985                            \\
ISSC    \>       IRAS Serendipitous Survey catalogue               \\
	\>        by Kleinmann et al., 1986.
\end{tabbing}
Some decametric sources with infrared emission are identified with
HII regions, and some with AGN (see Table 4).

Some of the UTR sources in Table~5 have also been found in
a cross-identification of the IRAS and TXS catalogues,
(cf.\ Trushkin and Verkhodanov, 1995, 1997). The full list of these
sources will appear in the Bulletin of the Special Astrophysical Observatory.

  We searched for the presence of any X-ray source registered in the
X-ray catalogues available in CATS within a circle  of radius $90\arcsec$
around the positions listed in Table~1. If such a source was found, an ``X''
is given in column~11 of Table~1.
   The following X-ray catalogues were searched by us:
\begin{tabbing}
1WGA    \quad\=    White et al., 1994     \\
EIN2S   \>         Moran et al., 1996     \\
EMSS    \>         Gioia et al., 1990      \\
	\>         Stocke et al., 1991    \\
RGN     \>         Neumann et al., 1994   \\
ROSAT   \>         Voges et al., 1994
\end{tabbing}

Tables~1 and 4 show that all X-ray identified UTR sourcs are well-known AGNs.


\clearpage
\mbox

\clearpage
\mbox{}
\clearpage
\mbox{}
\clearpage
\mbox{}
\clearpage
\mbox{}
\vspace{-0.9cm}
\section {Decametric Data for ``Cold'' Sources}
One of the UTR observational strips (Braude et al., 1979) overlaps
the ``Cold''
experiment survey carried out at RATAN-600 (Parijskij et al., 1991, 1992).

The results of a cross-identification with 18 ``Cold'' objects within
$60\arcsec$ from the positions in Table~1, together with the data from other
catalogues for these sources, are presented in Table~6.

Columns 1 and 3 of Table~6 give equatorial coordinates for B1950.0;
columns 2 and 4 list the corresponding errors, column 5 presents the
frequency in MHz, the flux density and its error (in Jy) are in
columns 6 and 7, and column 8 gives references to the catalogues. First we
give the name of a source in the RC and UTR catalogues.
For some UTR names we give the ID number next to the UTR name as
listed in Table 1.
In some cases the names from the 3C and 4C catalogues
are also presented, and stellar magnitudes~ (see also Table~4)~~ according
to the list VV83 (V\'eron-Cetty and V\'eron, 1983) are given.

The following references of catalogue names are used in Table 6:
\begin{tabbing}
11CMA  \quad\=   Wall and Peacock, 1985,              \\
11CMB  \>        F\"urst et al., 1990,                \\
21CMB  \>        Reich et al., 1990,                  \\
87GB   \>        Gregory and  Condon, 1991,           \\
87GBM  \>        Mingaliev and Khabrakhmanov, 1995,   \\
COLD1  \>        Parijskij et al., 1991,              \\
COLDB  \>        Bursov, 1996,                        \\
CUL    \>        Slee and Higgins, 1973,              \\
FSC2R  \>        Condon et al., 1995,                 \\
GB6    \>        Gregory et al., 1996,                \\
JVAS   \>        Patnaik et al., 1992,                \\
Ku79   \>        K\"uhr et al., 1979,                 \\
Ku81r  \>        K\"uhr et al., 1981,                 \\
MGVLA  \>        Lawrence et al., 1986,               \\
MITG1  \>        Bennett et al., 1986,                \\
MRC    \>        Large et al., 1991,                  \\
MSL    \>        Dixon, 1981,                         \\
NAIC   \>        Durdin et al., 1975,                  \\
NVSS   \>        Condon et al., 1998,                 \\
PKS05  \>        Drinkwater et al., 1997,             \\
PKS8G  \>        Wright et al., 1991,                 \\
PKS90  \>        Otrupcek and Write, 1990,            \\
PKSFL  \>        Duncan et al., 1993,                 \\
PKSW   \>        Wills, 1975,                         \\
PMN    \>        Wright et al., 1996,                 \\
POL\_T \>        Tabara and Inoue, 1980,              \\
QSOmm  \>        Reuter et al., 1997,                 \\
RCS86  \>        Roger et al., 1986,                  \\
RGB1   \>        Laurent-Muehleisen et al., 1997,     \\
RH74   \>        Readhead and Hewish, 1974,           \\
SRC\_N \>        Niell, 1972,                         \\
SoSRC  \>        White, 1992,                         \\
TXS    \>        Douglas et al., 1996,                \\
UCGC   \>        Cotton et al., 1999,                 \\
VLA4   \>        Perley, 1982,                        \\
VLAC   \>        Taylor, 1993,                        \\
VLAU3  \>        Ulvestad et al., 1981,               \\
VLBIS  \>        Preston et al., 1985,                \\
WB92   \>        White and Becker, 1992,              \\
Z2     \>        Amirkhanyan et al. 1989,             \\
Z2A    \>        Amirkhanyan et al., 1992,            \\
Z2\_95 \>        Gorshkov and Konnikova, 1995.        \\
\end{tabbing}

Radio spectra of these objects are shown in
Fig. 4.
We do not present in the figure spectral data for the whole radio
wavelength range, restricting ourselves to centimetre data, because not
for all the objects mm data are available.
Among the 18 ``Cold'' survey sources 7 objects have optical identifications.
One of these objects (RC\,J1148+0455 or GR\,1147+05) was investigated in
the ``Big Trio'' project and identified with a group of 8 galaxies having
stellar magnitudes of about $23\fm5$ (Parijskij et al., 2000).
The other six objects are identified with AGN.

\section {Ultra-steep spectrum UTR sources from FIRST}

  For the majority of UTR sources (97\%) identifications with
  NVSS sources are available (Condon et al., 1998)
(altogether 2253 identifications),
and for all UTR sources with $\delta>30\degr$ (1143 objects) we searched for
counterparts in the FIRST survey (White et al., 1997).
Among the FIRST and NVSS sources within 60$\arcsec$ from the positions in
Table~1, 552 from NVSS (with a beam size of $45\arcsec$) and 988 from
FIRST (with a beam size of $5\arcsec$) are resolved.
It would require more detailed observations to find out whether these
sources are merely complex in structure or consist of physically independent
components.

As mentioned earlier, the identifications with NVSS and FIRST have been used
to refine coordinates of the decametric sources. On the basis of these
coordinates, we extracted source samples for further investigation,
for instance, a sample of sources with ultra-steep spectra ($\alpha< -1.2$).

\section {Ultra-steep spectrum sources}

It is seen from the distribution of spectral indices (Fig. 1) that rather
a large proportion of objects have steep spectra at all of
the three frequencies.
To establish a list of sources which are candidates for distant objects,
we selected objects with ultra-steep ($\alpha\le-1.2$) and straight spectra,
which show an extent in the FIRST catalogue.

Generally, in the catalogue of 2314 radio counterparts to UTR sources
there are 422 S-type sources with ``very steep spectrum'' (VSS)
($\alpha\le-$1.0), and
for the present work we selected from these a subsample of 102
``ultra-steep spectrum'' (USS) objects ($\alpha\le-$1.2).
To further increase the radio-positional accuracy, we searched for
radio counterparts of USS sources in the February 1998 version of
the FIRST catalogue
(White et al.\ 1997), which resulted in 38 FIRST
counterparts for 23 UTR sources (see Table~7).
If an UTR source has more than one acceptable counterpart in FIRST,
we label these components with symbols ID1, ID2, ID3, etc.
Only one of the FIRST components (labeled GR1527+51\,(ID2)) is
truly unresolved by the FIRST beam of $\sim$\,5$''$
(i.e.\ it has the major and minor axes  $<$\,2$''$), while all other
objects have a multi-component or extended structure.
We checked the sources also in the lower resolution
NVSS at 1.4\,GHz (Condon et al.\ 1998). Usually, the greater
the source complexity, the larger the NVSS/FIRST flux ratio.
According to the NASA Extragalactic Database (NED), the
complex source GR0135$-$08
is identified with the $z=0.041$ galaxy MCG--02--05--020,
GR1214$-$03 is a LCRS QSO at $z=0.184$, and
GR1243+04 is a radio galaxy (4C+03.24) at $z=3.57$.
Our investigation of the Digital Sky Survey
(DSS~2) images, accessible via the Web-page
of the Space Telescope Science Institute (http://archive.stsci.edu/dss/)
for the unidentified
sources, is in progress.
\setcounter{table}{6}
{\small
\begin{table*}[!t]
\begin{center}
\caption{ The 38 FIRST counterparts to 23 USS sources ($\alpha\le-$1.2)
from the UTR catalogue. We list the UTR name, the size of the source complex
(or deconvolved size if a single component) from NVSS and FIRST, respectively,
the FIRST RA and Dec, the peak and integrated component
flux at 1.4\,GHz from FIRST,
their deconvolved major and minor axes, and major axis position
angle (N through E) from FIRST}
\begin{tabular}{|llrrcrrrrr|}
\hline
 UTR name       &$\alpha$& NVSS~& FIRST & RA ~(J2000)~ Dec  &  $S_p$~~ &  $S_i$~~ & Maj   &  Min &  PA\\
 ~~~(B1950)     &     & size ($''$) &  size ($''$) &  &  (mJy) & (mJy) & (\arcsec)~ &(\arcsec)~ & (\degr) \\
\hline
GR0002$+$00(ID2)&$-$1.52 & 16.  &  7.4 &000650.56$+$003648.4 &   47.6 &   75.9&   7.4 &   1.1 &  154 \\  
GR0135$-$08     &$-$1.34 & 28.  & 20.~~&013714.87$-$091155.4 &    5.2 &   44.3&  30.8 &   7.5 &   98 \\  
		&       &      &      &013715.08$-$091203.3 &   10.2 &   36.9&  15.6 &   4.3 &   90 \\  %
		&       &      &      &013715.45$-$091155.8 &    8.2 &   25.7&  13.0 &   5.4 &  148 \\  %
		&       &      &      &013716.22$-$091149.5 &   10.0 &   39.3&  11.3 &   9.3 &  119 \\  %
GR0257$-$08     &$-$1.20 &$<$18.&  4.5 &025919.15$-$074501.2 &  162.5 &  211.9&   4.5 &   1.3 &   59 \\  
GR0257$-$08(ID2)&$-$1.23 & 80.  & 83.~~&030040.22$-$075302.2 &   18.4 &   52.8&  15.6 &   3.0 &  174 \\  
		&       &      &      &030040.56$-$075259.6 &   10.1 &  122.8&  30.7 &  13.1 &  163 \\  %
		&       &      &      &030042.99$-$075413.8 &    7.0 &  34.2 &  16.2 &   8.4 &   32 \\
		&       &      &      &030042.99$-$075358.0 &   13.6 &  47.8 &  12.3 &   7.4 &  152 \\
		&       &      &      &030043.60$-$075418.3 &    6.4 &  22.0 &  11.2 &   7.6 &   75 \\
		&       &      &      &030043.75$-$075407.2 &    6.6 &  22.4 &  15.6 &   4.8 &  170 \\
GR0723$+$48(ID2)&$-$1.26 & 47.  &  3.1 &072651.18$+$474041.5 &   23.8 &   29.1&   3.1 &   2.0 &  175 \\  
		&       &      &      &072655.00$+$474051.0 &   51.8 &   75.7&   5.6 &   1.1 &   85 \\
GR0818$+$18     &$-$1.28 & 24.  & 22.~~&082032.48$+$192731.3 &   31.1 &   49.2&   6.1 &   1.7 &  174 \\  
		&       &      &      &082032.73$+$192709.0 &   76.4 &   98.5&   3.9 &   1.6 &    8 \\  %
GR0858$-$02(ID2)&$-$1.54 & 70.  & 55.~~&085935.06$-$015842.1 &   11.7 &   27.2&  10.2 &   3.2 &  122 \\  
		&       &      &      &085936.10$-$015851.8 &    9.8 &   15.9&   6.2 &   2.8 &  128 \\  %
		&       &      &      &085938.21$-$015908.1 &   20.2 &   43.6&   8.1 &   4.5 &  133 \\  %
GR0910$+$48     &$-$1.22 & 36.  & 29.~~&091359.00$+$482738.0 &   26.9 &   88.6&  11.6 &   5.2 &  100 \\  
		&       &      &      &091401.83$+$482729.2 &   39.6 &  110.2&  10.9 &   3.9 &  106 \\  %
GR0922$+$42(ID2)&$-$1.21 &$<$19.&  2.7 &092559.66$+$420335.3 &  199.7 &  244.1&   2.7 &   2.4 &   98 \\  
GR0942$+$54     &$-$1.24 & 14.  &  8.~~&094618.12$+$543003.8 &   51.1 &   57.1&   2.5 &   0.9 &   36 \\  
		&       &      &      &094618.53$+$543010.1 &   75.3 &   80.6&   1.7 &   1.1 &   16 \\  %
GR1149$+$42     &$-$1.24 &$<$17.&  5.0 &115213.58$+$415344.9 &   83.7 &  115.4&   5.0 &   1.0 &   18 \\  
GR1214$-$03     &$-$1.26 &~~70.\,?&3.2 &121755.30$-$033722.0 &  176.9 &  208.9&   3.2 &   1.2 &  111 \\  
GR1223$-$00     &$-$1.79 & 25.  & 25.~~&122722.97$-$000813.8 &    5.8 &   13.6&   8.4 &   5.0 &  100 \\  
		&       &      &      &122724.54$-$000821.1 &    8.4 &   16.9&   7.1 &   4.6 &  116 \\  %
GR1243$$+04     &$-$1.25 &$<$18.&  6.8 &124538.38$+$032320.1 &  249.4 &  379.9&   6.8 &   1.3 &  158 \\  
GR1318$+$54     &       & 25.  & 23.~~&132202.81$+$545758.1 &   30.3 &   90.3&   9.5 &   5.9 &   82 \\  
		&       &      &      &132205.33$+$545805.3 &    8.6 &   35.9&  10.7 &   8.6 &   55 \\  %
GR1320$+$43     &$-$1.32 &$<$19.&  3.3 &132232.32$+$425726.5 &  129.3 &  154.6&   3.3 &   1.0 &   81 \\  
GR1355$+$01(ID3)&$-$1.26 &$<$19.& 10.~~&135821.64$+$011442.0 &   77.0 &   88.4&   2.9 &   1.4 &   42 \\  
		&       &      &      &135822.08$+$011449.4 &  169.9 &  180.9&   1.6 &   1.3 &   32 \\  %
GR1447$+$57     &$-$1.26 &$<$19.&  4.5 &144630.04$+$565146.8 &   75.2 &   99.2&   4.5 &   0.7 &  148 \\  
GR1527$+$51(ID2)&$-$1.22 &$<$19.&  1.4 &152828.36$+$513401.4 &  203.9 &  212.8& $<$2. & $<$2. &  140 \\  
GR1539$+$53(ID2)&$-$1.24 &$<$18.&  6.4 &154144.69$+$525054.5 &   87.0 &  136.7&   6.4 &   0.8 &   66 \\  
GR1613$+$49     &$-$1.24 &$<$19.&  3.0 &161631.16$+$491908.2 &   56.1 &   66.2&   3.0 &   1.3 &   10 \\  
GR1731$+$43     &$-$1.56 & 42.  & 40.~~&173333.87$+$434318.6 &   32.9 &   57.9&   6.3 &   3.0 &  164 \\  
		&       &      &      &173334.26$+$434300.8 &   22.8 &   27.3&   3.1 &   1.5 &  173 \\  %
		&       &      &      &173334.41$+$434251.4 &   12.3 &   15.8&   3.3 &   2.4 &  159 \\  %
		&       &      &      &173334.58$+$434239.8 &   27.3 &   51.0&   7.5 &   2.3 &  150 \\  %
GR2211$-$08(ID2)&$-$1.39 &$<$19.&  8.6 &221519.65$-$090005.8 &   75.2 &  125.0&   8.6 &   1.4 &  176 \\  
\hline
\end{tabular}
\end{center}
\end{table*}
}

\section {Conclusion}

   We have shown that our iterative method of radio identification and
cross-identification with existing catalogue material serves as a first step
for an optical identification
of radio sources observed with a large
antenna beam.

We present a catalogue of 2316 radio counterparts to 1822 UTR sources.
This is due to the fact that 395 UTR sources
have
more than one counterpart contributing significantly to the decametric
flux. We found radio counterparts at non-decametric frequencies for
all but three UTR sources,
compared to a fraction of 93\% radio-identified
sources in the original UTR publication and 89\% in the electronic version of
UTR.

Through a cross-identification with optical catalogues we increased the optical
identification rate from 19\% in the original publications of the UTR
catalogue to presently 32\%.

We also mention the presence of likely counterparts for UTR sources in the
infrared and X-ray regimes, and construct a subsample of 38 ultra-steep
spectrum sources having counterparts in the FIRST catalogue.

\begin{acknowledgements}
The authors are thankful
to our colleagues Miroshnichenko A.P. and Krivitskij D.
from the Institute of Radio Astronomy (Kharkov) for providing data and
useful discussion, Sergej Trushkin and Vladimir Chernenkov (SAO RAS),
co-creators of the CATS database, for useful discussion.
H.A. thanks CONACyT for financial support under grant 27602-E.
Thanks are also due to N.F.\,Vojkhanskaya for valuable comments on reading
the manuscript.
\end{acknowledgements}

\end{document}


\special{landscape}


{\tiny

\begin{center}
\vspace{0.5cm}

\topcaption{Multifrequency data for 18 RC radio sources}
\tablefirsthead{\hline
		    \multicolumn{1}{|c}{$\alpha$}
		 &  \multicolumn{1}{c}{$\sigma_\alpha$}
		 &  \multicolumn{1}{c}{$\delta$}
		 &  \multicolumn{1}{c}{$\sigma_\delta$}
		 &  \multicolumn{1}{c}{$\nu$}
		 &  \multicolumn{1}{c}{S}
		 &  \multicolumn{1}{c}{$\sigma_S$}
		 &  \multicolumn{1}{c|}{Catalogue}        \\
\hline
		    \multicolumn{1}{|c}{$~~^h~~^m~~^s$}
		 &  \multicolumn{1}{c}{$~~^s$}
		 &  \multicolumn{1}{c}{~~\degr~\arcmin~~\arcsec}
		 &  \multicolumn{1}{c}{~~\arcsec}
		 &  \multicolumn{1}{c}{MHz}
		 &  \multicolumn{1}{c}{Jy}
		 &  \multicolumn{1}{c}{Jy}
		 &  \multicolumn{1}{c|}{}            \\
\hline
		    \multicolumn{1}{|c}{1}
		 &  \multicolumn{1}{c}{2}
		 &  \multicolumn{1}{c}{3}
		 &  \multicolumn{1}{c}{4}
		 &  \multicolumn{1}{c}{5}
		 &  \multicolumn{1}{c}{6}
		 &  \multicolumn{1}{c}{7}
		 &  \multicolumn{1}{c|}{8}           \\
 \hline              }
\tablehead{ \hline
		    \multicolumn{1}{|c}{$~~^h~~^m~~^s$}
		 &  \multicolumn{1}{c}{$~~^s$}
		 &  \multicolumn{1}{c}{~~\degr~\arcmin~~\arcsec}
		 &  \multicolumn{1}{c}{~~\arcsec}
		 &  \multicolumn{1}{c}{MHz}
		 &  \multicolumn{1}{c}{Jy}
		 &  \multicolumn{1}{c}{Jy}
		 &  \multicolumn{1}{c|}{}            \\
		    \multicolumn{1}{|c}{1}
		 &  \multicolumn{1}{c}{2}
		 &  \multicolumn{1}{c}{3}
		 &  \multicolumn{1}{c}{4}
		 &  \multicolumn{1}{c}{5}
		 &  \multicolumn{1}{c}{6}
		 &  \multicolumn{1}{c}{7}
		 &  \multicolumn{1}{c|}{8}            \\
     \hline          }
\tabletail{\hline}
\begin{supertabular}{|lllrrlll|}
	     &       &             &      &      &       &         &       \\
\multicolumn{8}{|l|}
	{{\bf RC J0038+0449}, GR 0037+05(ID2)} \\
00 36 59     &    76 & 05 57 36    &      &  12.6 & $<$288 &       & UTR   \\
00 37 30     &    22 & 05 00 36    &      &  16.7 &    59  &    32 & UTR   \\
00 36 40     &    37 & 04 52 48    &      &    25 &    17  &    70 & UTR   \\
00 36 00.043 &  0.06 & 04 34 21.54 & 0.48 &   365 &  0.838 & 0.061 & TXS   \\
00 36 00.2   &       & 04 34 39    &      &   408 &   0.75 &   .06 & MRC   \\
00 35 58.7   &   4.0 & 04 34 14    &   78 &   611 &  0.627 &  .103 & NAIC  \\
\hline
\end{supertabular}
\end{center}